\documentclass{sig-alternate-2013}

\usepackage[tight,footnotesize]{subfigure}
\usepackage{cite}
\usepackage{multirow}
\usepackage{framed}
\usepackage{amsmath}
\usepackage{amssymb,amsfonts}
\usepackage{threeparttable}
\usepackage{algorithm} %format of the algorithm
\usepackage{algorithmic} %format of the algorithm
\usepackage{color}% use color
\usepackage{caption}% use caption
\usepackage{colortbl} % use table color
\usepackage{framed}
\definecolor{shadecolor}{rgb}{0.9,0.9,0.9}

\usepackage{times}

\newfont{\mycrnotice}{ptmr8t at 7pt}
\newfont{\myconfname}{ptmri8t at 7pt}
\let\crnotice\mycrnotice%
\let\confname\myconfname%

\permission{Permission to make digital or hard copies of all or part of this work for personal or classroom use is granted without fee provided that copies are not made or distributed for profit or commercial advantage and that copies bear this notice and the full citation on the first page. Copyrights for components of this work owned by others than ACM must be honored. Abstracting with credit is permitted. To copy otherwise, or republish, to post on servers or to
redistribute to lists, requires prior specific permission and/or a fee. Request permissions from Permissions@acm.org.}
\conferenceinfo{SIGIR'16,}{July 17-21, 2016, Pisa, Italy.}
\copyrightetc{\copyright~2016  ACM. ISBN \the\acmcopyr}
\crdata{978-1-4503-4069-4/16/07\ ...\$15.00.\\
DOI: http://dx.doi.org/10.1145/2911451.2914744}

\newcommand{\nop}[1]{}

\usepackage[usenames,dvipsnames,svgnames,table]{xcolor}

\begin{document}

\title{Community-based Cyberreading for Information Understanding}
%\subtitle{How to Help Readers Better Understanding the Scientific Publications via Physical or  Virtual  Collaboration?}

\numberofauthors{4} %  in this sample file, there are a *total*
\author{
% 1st. author
\alignauthor
Zhuoren Jiang\\
      \affaddr{Institute of Computer Science and Technology}\\
      \affaddr{Peking University}\\
      \affaddr{Beijing, China, 100871}\\
       \email{jiangzr@pku.edu.cn}
% 2nd. author
 \alignauthor
Xiaozhong Liu\titlenote{Xiaozhong Liu and Liangcai Gao are the corresponding authors}\\
       \affaddr{School of Informatics and Computing}\\
       \affaddr{Indiana University Bloomington}\\
       \affaddr{Bloomington, IN, USA, 47405}\\
       \email{~liu237@indiana.edu}
% 3rd. author
\alignauthor
Liangcai Gao\titlenote{}\\
       \affaddr{Institute of Computer Science and Technology}\\
       \affaddr{Peking University}\\
       \affaddr{Beijing, China, 100871}\\
       \email{glc@pku.edu.cn}
% 3rd. author
\alignauthor
Zhi Tang\\
       \affaddr{Institute of Computer Science and Technology}\\
       \affaddr{Peking University}\\
       \affaddr{Beijing, China, 100871}\\
       \email{tangzhi@pku.edu.cn}
}

\maketitle
\begin{abstract}

Although the content in scientific publications is increasingly challenging, it is necessary to investigate another important problem, that of scientific information understanding. For this proposed problem, we investigate novel methods to assist scholars (readers) to better understand scientific publications by enabling physical and virtual collaboration. For physical collaboration, an algorithm will group readers together based on their profiles and reading behavior, and will enable the cyberreading collaboration within a online reading group. For virtual collaboration, instead of pushing readers to communicate with others, we cluster readers based on their estimated information needs. For each cluster, a learning to rank model will be generated to recommend readers' communitized resources (i.e., videos, slides, and wikis) to help them understand the target publication. 

\end{abstract}

%\category{H.3.3}{Information Storage and Retrieval}{Information Search and Retrieval}
%\nop{
%\terms{System, Algorithms, Experimentation, Measurement}
%}
% NOT required for Proceedings

\vspace*{-0.75\baselineskip}\section{INTRODUCTION AND MOTIVATION}\label{sec:intro}

STEM (Science, Technology, Engineering, and Mathematics) publications, for various reasons, generally do not place a premium on writing for readability, and young scholars struggle to understand the scholarly literature available to them. Unfortunately, few efforts have been made to help graduate students and other junior scholars understand and consume the essence of those scientific readings. In other words, while scholarly search and recommendation algorithms are well-documented, few studies address a critical problem -- scientific information understanding. Compared to search and recommendation, information understanding can be more challenging. It can be more difficult to characterize user (implicit and explicit) information needs when a user is reading a paper, and it can be even more challenging to address the personalization problem. For example, one can assume that when a user is reading a paper, he/she may require different kinds of assistance to understand the specific sentences/paragraphs/sections/formulas in the paper. Meanwhile, given the same piece of text/formula in a paper, other users may need different types of help, such as personalized information understanding.

In this study, we propose two different kinds of information understanding scenarios: \textbf{information understanding via reader-reader collaboration} and \textbf{information understanding via Open Education Resources (OER) recommendation}. The former scenario is based on prior studies in collaborative scaffolding (a.k.a. peer support) studies in education domain \cite{pea2004social, novak2012educational}. Despite some criticism, evidence shows that the benefits of this type of peer support exceed its limitations, and that readers can learn from each other for information understanding (physical collaboration). The latter type is based on the pilot-evidence that accessing multi-modal OER about a scholarly publication, including presentation videos, slides, source code or Wikipedia pages, in a collaborative framework will enhance a student's ability to understand the paper itself~\cite{liu2015scientific}. However, different kinds of readers, given the same publication content, may prefer different kinds of OERs for information understanding. A reader community (sharing the similar information needs) may prefer similar OERs, and when we don't have reader statistics, (virtual) `community' can be a latent variable.  

%cite -> A case study of using a social annotation tool to support collaboratively learning 

In this paper, we propose a novel collaborative cyberreading environment, OER-based Collaborative PDF Reader (OCPR), which enables reader collaboration and community-based OER recommendation for information understanding. 

The contribution of this paper is threefold. First, we propose a novel method to group readers into a number of physical communities, and readers from the same groups are more likely to collaborate while reading. Second, when readers don't want to collaborate, we propose the new algorithm to group similar users into virtual communities for information need estimation and OER recommendation. Last but not least, an experiment (with 60 participants) is employed to validate the usefulness of the proposed methods. Experiment results show our methods can significantly improve OER recommendation performance for information understanding.
%\vspace*{-0.02\baselineskip}

\vspace*{-0.55\baselineskip}
\begin{figure*}[t]
  \centering
  \includegraphics[width=0.92\textwidth]{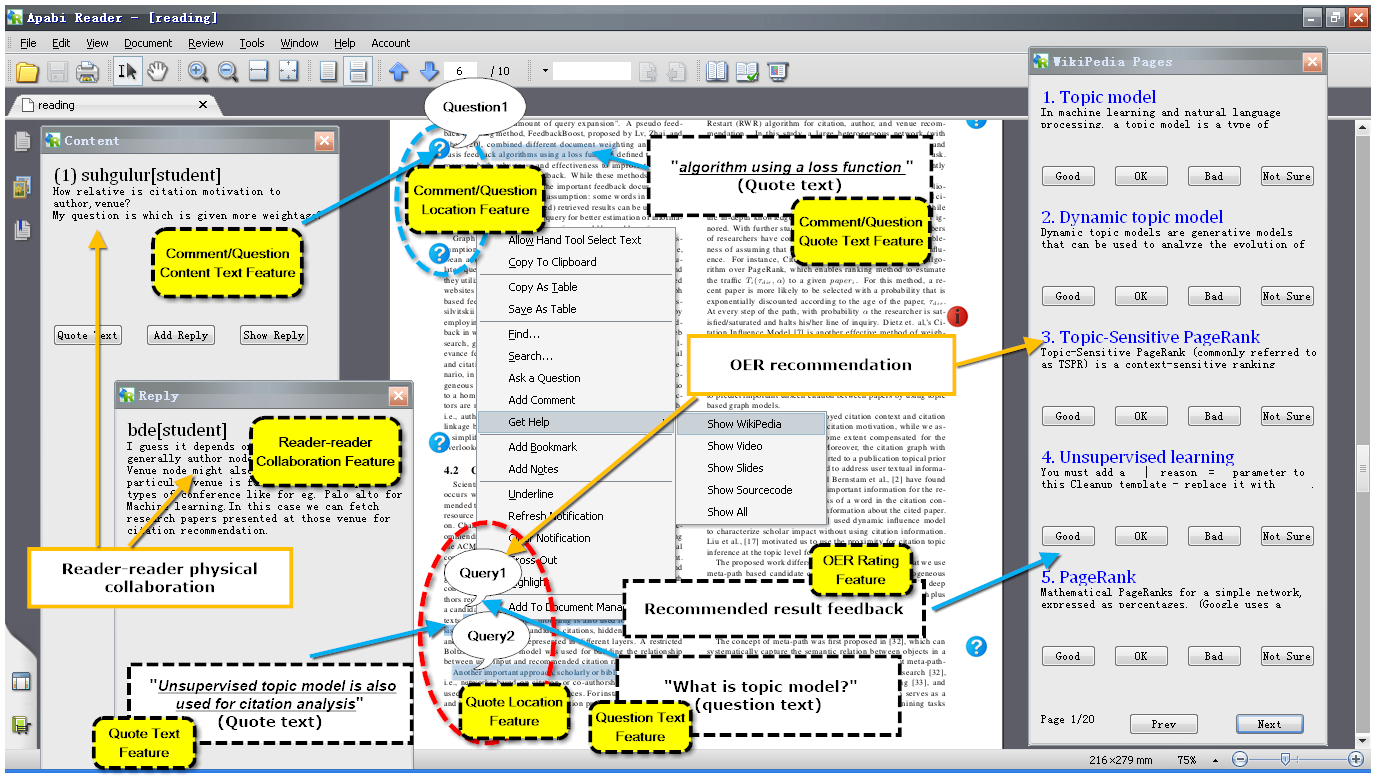}
  \setlength{\abovecaptionskip}{1pt}
  \setlength{\belowcaptionskip}{1pt}
  \caption{OER-based Collaborative PDF Reader System.}
  \label{fig:system}
\end{figure*}
\vspace*{-0.55\baselineskip}

\vspace*{-0.55\baselineskip} \section{RELATED WORK}

From an educational viewpoint, scaffolding has been widely used in educational research~\cite{pea2004social}. In particular, a collaborative scaffolding study \cite{johnson2010individual} finds that students achieved better reading comprehension and metacognitive skills when reading within the context of small team collaboration, e.g., comment on part of a document, share ideas, and provide feedback among a group of users \cite{su2010web}. 

However, prior studies also found collaborative scaffolding approaches can be quite limited~\cite{novak2012educational}, and readers may not have enough incentives to collaborate while reading. Not until recently did researchers begin to focus on the usefulness of OER. For instance, Dennis, et al.~\cite{dennisimproving} found that video presentations had significantly positive impacts on students' learning based on quiz scores. More recently, Liu~\cite{liu2013generating, liu2013answering} found OERs, such as YouTube video, technology slides, and Wikipedia pages, can help students better understand scientific readings. Meanwhile, information retrieval and text mining methods, i.e., language model and BM25, had been proven to be useful for building OER index for a large number of publications automatically \cite{liu2013generating}. However, this approach cannot accurately address students' emerging information need while reading a paper, especially for personalized or communitized reader information need estimation.  
\vspace*{-0.85\baselineskip}

\vspace*{-0.75\baselineskip}\section{RESEARCH METHODS}

%In this section we discuss the research method in detail which includes: constructing an OCPR reading system for reading feature collection~(\ref{sec:feature}), enabling physical readers' collaboration~(\ref{sec:physical}), and generating a virtual community for OER recommendation~(\ref{sec:virtual}).

\vspace*{-0.75\baselineskip}\subsection{Reader Profile and Reading Behavior}\label{sec:feature}

To generate physical and virtual communities and assist readers in understanding publications, we need to generate a number of features for reader clustering. There are two kinds of features: reader profile features (RPF) and reading behavior features (RBF). Note that RPF can be important but can be missing in some cases (e.g., online journal or proceeding paper readers). Compared to RPF, RBF can be more reliable. 

To capture RBF and implement information understanding, we designed OCPR, which is depicted in Figure \ref{fig:system}. OCPR enables two types of scaffoldings:

\textbf{Collaborative Scaffolding}: enabling readers to asynchronously annotate PDF publications or provide feedback and comments to existing information needs from other readers. By using the OCPR, any reader can view other readers' information needs, such as highlighted (quoted) parts of PDFs or specific questions/comments (as shown in Figure \ref{fig:system}, the question or exclamation marks on the side of PDF). These readers also can provide their own replies and feedbacks to the existing or proposed information needs (reader collaboration). 

\textbf{OER-based Scaffolding}: capturing evidence of students' emerging implicit or explicit information needs when reading a scientific paper and recommending high quality OERs to address their information needs. As Figure 1 shows, by using OCPR, readers can ask a specific question given a piece of text which serves as evidence of an explicit information need or highlight (quote) part of a text in the paper as evidence of an implicit information need. In either case, the OCPR is able to recommend OERs given the textual content and the target paper, i.e., $P(OER|text,paper)$.

The detailed RPF and RBF features are listed in Table \ref{tab:feature}. 

\begin{table*}[t] 
\scriptsize
\centering
 \setlength{\abovecaptionskip}{1pt}
 \setlength{\belowcaptionskip}{1pt}
\caption{All features for information understanding}
\label{tab:feature}
\begin{tabular}{|l|l|m{12cm}|}
\hline
\textbf{Feature Group} & \textbf{Feature Name}         & \textbf{Feature Description}                                     \\ \hline
RPF &  Reader Profile  &  \emph{Feature for explicit reader characterization, i.e., reader research topics/expertises (for scholars) or reader course taking history (for students). This kind of features is not always available.}  \\ \hline
\multirow{8}{*}{RBF} 
&  Quote Location  &       \emph{Feature for location of reader's quote text, we assume that if readers share the similar information needs, they will launch queries at the similar place of the reading. To generate this kind of feature, we first cluster all of the readers' queries by their quote text location coordinates, then we use the query count of each reader in each cluster as the feature value.}                            \\     \cline{2-3}
&  Quote Text     &  \emph{Feature for quote text, we assume that if readers share the similar information needs, their quote texts(implicit information need) will have a higher text similarity. We use term frequency of reader's quote text as this kind of feature.}  \\ \cline{2-3}
&  Question Text     &   \emph{Feature for question text, we assume that if readers share the similar information needs, their question texts(explicit information need) will  have a higher text similarity. We use term frequency of reader's question text as this kind of feature. } \\ \cline{2-3}
& OER Rating   &  \emph{Feature for reader's OER ratings,  we assume that if readers share the similar information needs, their preference of OER will be similar. We use OER ratings of readers as this kind of feature.} \\ \cline{2-3}
&  Comment/Question Location     & \emph{Feature for location of reader's quote text of comment/question, we assume that if readers share the similar information needs, they will comment or have a question at the similar place of the reading. To generate this kind of feature, we first cluster all the readers' comments/questions by their quote text location coordinates, then we use the comment/question count of each reader in each cluster as the feature value.}   \\ \cline{2-3}
&  Comment/Question  Quote Text      &   \emph{Feature for quote text of reader's comment/question, we assume that if readers share the similar information needs, their quote texts of comments/questions will have a higher text similarity. We use term frequency of reader's quote text of comments/questions as this kind of feature.}  \\ \cline{2-3}
&  Comment/Question Content Text     &   \emph{Feature for content text of reader's comment/question, we assume that if readers share the similar information needs, their content texts of comments/questions will  have a higher text similarity. We use term frequency of reader's content text of comment/question as this kind of feature.} \\ \cline{2-3}
&  Reader-reader Collaboration &  \emph{Feature for reader's communication behavior, we assume that if readers share the similar information needs, they will communicate more frequently. We use reader's reply relation as this kind of feature.} \\ \hline
\end{tabular}
\end{table*}

\vspace*{-0.55\baselineskip}\subsection{To Enable Physical Collaboration }\label{sec:physical}

For any publication, a large number of readers could potentially access and read it. For this approach, the proposed task is to predict reader physical collaboration. In other words, by using the features presented in Table 1 (except for `reader-reader collaboration'), we will cluster users together, and inner cluster users are more likely to collaborate with each other. This information can be important for online cyberreading environment, e.g., OCPR. For instance, when a large number of readers are reading a paper, $user_x$'s question(s) will be broadcast to the target reader group ($user_x$ belongs to) for potential collaboration. Compared to most existing clustering problems, reader cluster copes with a smaller number of instances, so we chose the K-medoids clustering algorithm \cite{gullo2008clustering}. Compared to K-means and EM clustering, K-medoids is more efficient with small dataset. For evaluation, we use ``reader-reader collaboration'' data extracted by OCPR system. 

However, incentive can be a challenge for physical collaboration. Prior studies in scaffolding show that even though reader collaboration may be potentially useful, not all the readers would like to participate the online reading collaboration (e.g., reply to communicate with other readers). 

%As shown in Figure~\ref{fig:system}, reader can reply other readers' comments and questions, we consider that is reader-reader physical collaboration. We assume that if we find out ``who will be more likely to reply the target students'', we can make physical collaboration between readers easier and more efficient. In order to achieve this goal, we are trying to create communities in the readers, here, ``community'' of readers means the readers in the same community have interests in common, so readers in the same community are more likely to collaborate while they are less likely to collaborate if they are from different communities.

%For the community generation, we use cluster algorithm based on different feature group described in section~(\ref{sec:feature}), then we compare the clustering performance(Precision, Recall, F1-measure) to find out the best community divisions. 

\vspace*{-0.55\baselineskip}\subsection{To Enable Virtual Collaboration }\label{sec:virtual}

To address the limitations of the physical collaboration, we propose a two-step method for OER-based scaffolding. First, by using the features in Table 1, we group the readers based on RPF and/or RBF. The readers from the same group are more likely to share the similar information needs. Second, for each group, we train an OER-recommendation (ranking) model, and the communitized recommendation model can recommend the optimized OERs, which could highly likely help readers (from the same group) to understand the target publication. Note that by using this method, readers from the same group don't necessarily know each other or need to physically collaborate with others. Instead, the group information can be used to train the OER-recommendation models. For instance, given the same piece of text in a paper, some CS students may prefer to access the source code for information understanding, while another group (e.g., information science students) may like to watch a video or presentation slides. 

%%%%%%%%%%%You need to propose your method here! Why use Max Entropy????????? `In this section, we propose a new method to optimize the clustering performance based on the observation that the RPF quality is higher than RBF but RPF features are more expensive...      '
As mentioned earlier, RPF quality can be higher than RBF (as RBF can be noisy), but RPF is not always available. Motivated by these observations, we propose a new method to optimize the clustering performance: (1) select the readers whose RPF are available; (2) cluster this part of readers using K-medoids algorithm based on RPF; (3) use the clustering result for a reader as his/her community label and train a Maximum Entropy classifier \cite{nigam1999using} based on readers' RBF; (4) use trained classifier to predict the community of the reader who only has RBF. Unlike clustering, the proposed method can enhance the informative RBF while punishing noisy RBF. 

The Maximum Entropy classifier is a discriminative classifier commonly used in NLP and IR problems. Unlike the other similar probabilistic classifiers, the Max Entropy does not assume that the features are conditionally independent of each other. \nop{Moreover, Maximum Entropy is used when we can't assume the conditional independence of the features.}This is particularly true in this study where our features are usually not independent (i.e., readers with same background tend to have similar reading behavior). %Due to the minimum assumptions that the Maximum Entropy classifier make, we use it when we don't know anything about the prior distributions and when it is unsafe to make any such assumptions. 

In this study, by using the meta-search algorithm presented in \cite{liu2013generating}, we indexed four kinds of OERs for information understanding, such as Wikipedia pages, presentation slides, presentation videos, and source code, from various sites (e.g., Videolecture, and GitHub).

As shown in Figure~\ref{fig:system}, readers can get OER recommendation from the system when they are reading a scientific paper and highlighting a piece of text in the paper. Readers can also provide judgments for the recommended OERs, e.g., rated as `Good,' `OK,'  and `Bad.' We then accumulate all the OER judgments from the same reader group to train a learning to rank model for OER recommendation. For learning to rank model generation, we employed 56 text- and graph-based ranking features\footnote{\scriptsize The detailed feature list is available in the project website: http://59.108.48.32:8086/CCIU/algorithm.html}. For text ranking method, we recommend OER based on the language model between the query text (in the paper) and OER content. For graph-based features, we created a heterogeneous graph, and the OER recommendation is conceptualized as a random walk problem on a heterogeneous graph. Three kinds of vertices -- paper, topic and OER -- are interconnected by using different kinds of edges. We created different kinds of meta-paths to address different ranking hypotheses, and each meta-path carries one or more types of ranking information \cite{Sun+11}. For instance, $K^{*}\overset{s}{\rightarrow} R\overset{r}{\leftarrow} R_{video}^{?} $ is a meta-path function, and it means the candidate OER should relate to the important topic's related OER and candidate OER's type is video. We use the random walk probability from starting (query) vertices to the target OER vertices following a meta-path: $\scriptsize r(v_i^{(1)},v_j^{(l+1)}) = \sum_{t= v_i^{(1)}\rightsquigarrow v_j^{(l+1)}} RW(t)$, where $t$ is a tour from $v_i^{(1)}$ to $v_j^{(l+1)}$ following the meta-path, and $RW(t)$ is the simulated random walk probability of the tour $t$.

As this study is not focusing on learning to rank, we used a relative simple algorithm (Coordinate Ascent~\cite{L2R}, which iteratively optimizes a multivariate objective ranking function) for OER ranking feature integration and algorithm evaluation.

%this table is for experiment(for output purpose)
\vspace*{-0.55\baselineskip}
\begin{table*}[t]
\scriptsize
\centering
 \setlength{\abovecaptionskip}{1pt}
 \setlength{\belowcaptionskip}{2pt}
\caption{Virtual collaboration performance with clustering result (* means the best result)}
\label{tab:vcp}
\begin{threeparttable}
\begin{tabular}{| l | l | l | l | l | l | l | l |}
\hline
Feature Name  & MAP@3 & MAP@5 & MAP@all  & nDCG@3 & nDCG@5 & nDCG@all & MRR \\ \hline
Globe Learning-to-rank(baseline) &  0.5582 & 0.6677 & 0.7012 & 0.6346 & 0.6667 & 0.6822 & 0.7340 \\ \hline
RPF-C & 0.5643 &	0.6545 &  0.6958 & 0.6247 & 0.6546 & 0.6781 & 0.7257  \\ \hline
RPF-TB   &  0.5699	& 0.6800	& 0.7070	& 0.6356	& 0.6708	& 0.6843 & 0.7357 \\ \hline
RPF-all &   \textbf{0.5849\tnote{*}} &  \textbf{0.6960} & \textbf{0.7148} &  \textbf{0.6490\tnote{*}} &  \textbf{0.6837} &  \textbf{0.6938} &  \textbf{0.7399}  \\ \hline
RBF & 0.5523 & 0.6632 & 0.6938	 & 0.6270 & 0.6625 & 0.6755 & 0.7209 \\ \hline
RPF-all+RBF & 0.5547 & 0.6561 & 0.6868 & 0.6168 & 0.6491 & 0.6648 & 0.7147 \\ \hline
%RBF with Balanced Winnow & 0.4700 & 0.5932  & 0.6080 & 0.5156 & 0.5914 & 0.6004 & 0.6357 \\ \hline
%RBF with C4.5 & 0.4886 & 0.5943 & 0.6367	 & 0.5679 & 0.6031 & 0.6224 & 0.6650 \\ \hline
RBF with Max Entropy &\textbf{0.5818} & \textbf{0.6985\tnote{*}} &  \textbf{0.7155\tnote{*}} & \textbf{0.6481} & \textbf{0.6881\tnote{*}} & \textbf{0.6966\tnote{*}} & \textbf{0.7401\tnote{*}} \\ \hline
%RBF with Naive Bayes & 0.5238 &0.6295 & 0.6649 & 0.5951 & 0.6316 & 0.6471 & 0.6947 \\ \hline
\end{tabular}
\end{threeparttable}
\end{table*}
\vspace*{-0.3\baselineskip}

\vspace*{-0.25\baselineskip}
\section{EXPERIMENT}

\vspace*{-0.65\baselineskip}\subsection{Data}
We tested our collaboration algorithms in a real learning environment. A graduate-level information retrieval course at Indiana University is used for this experiment. Sixty students (masters and PhDs) voluntarily participated in this experiment, and they were required to use the OCPR system for a semester~\footnote{\scriptsize The dataset can be download in the project website: http://59.108.48.32:8086/CCIU}. They could use OCPR functions to read scientific readings, ask questions, write comments and receive access to the system-recommended OERs. Meanwhile, we asked each participant to provide OER relevance judgments for the top five system-recommended OERs. There are a total of 39 valid users (who launched requests in more than six different readings and has made more than eight OER ratings) for this experiment. Meanwhile, there are total 626 OER recommendation requests with 3,551 valid OER usefulness judgments. For reader collaboration, there are total 568 messages sent among 39 users. 

In this experiment, we collect two kinds of RPF: 1. students' course taking history (`RPF-C' boolean features), e.g., if student has taken \textit{`machine learning'} and \textit{`statistics'} courses. 2. students' technical background survey (`RPF-TB' ordinal features from 1 to 4), e.g., if they have expertise in \textit{`R'} and \textit{`NoSQL'}. 

At the backend of OCPR, we created text and graph indexes for OER recommendation. For paper, we used 41,370 publications from 1,553 venues (mainly from the ACM digital library). The paper vertices are connected to 9,263 keyword labeled topics. By using meta-search, we collected a total of 1,112,718 OERs.  %\jeff{"As aforementioned", actually we didn't mention this message in our paper, should we add more detailed information?} there are \jeff{three} kinds of vertices and 10 kinds of relations in this heterogeneous graph. 

\vspace*{-0.65\baselineskip}\subsection{Experiment results}
\vspace*{-0.1\baselineskip}

As Table \ref{tab:pcp} shows, we use different kinds of feature sets to cluster the students, and the evaluation metrics are precision, recall, and F1. In the experiment, we set the cluster number as 3 (for total 39 participants). The ground truth is student-student communication data via OCPR. For instance, if $student_1$ and $student_2$ communicate via OCPR, they should belong to the same cluster. Otherwise, they should belong to different clusters. Evaluation results shows that \textbf{RPF-all} achieves the best recall, while \textbf{RPF-TB} optimizes the precision. When RPF features are not available, RBF features achieve 0.3862 F1, which is lower than RPF. 

\nop{We evaluate the community division performance based on the reply relation between students. The community division based on Programming Skill Feature can achieve the best Precision and F1-measure while Course Taken Feature and RPF can achieve the best Recall. } 

\vspace*{-0.2\baselineskip}
\begin{table}[!htbp]
\scriptsize
\centering
 \setlength{\abovecaptionskip}{2pt}
 \setlength{\belowcaptionskip}{2pt}
\caption{Physical collaboration performance}
\label{tab:pcp}
\begin{threeparttable}
\begin{tabular}{| l | l | l | l |}
\hline
Feature Name  & Precision & Recall & F1-measure  \\ \hline
RPF-C &  0.3536 &  \textbf{0.4502\tnote{*}} & 0.3961  \\ \hline
RPF-TB  &  \textbf{0.4479\tnote{*}} & 0.4280 & \textbf{0.4377\tnote{*}}  \\ \hline
RPF-all &  0.3567 & \textbf{0.4502\tnote{*}} & 0.3980  \\ \hline
RBF & 0.3625 & 0.4133 & 0.3862 \\ \hline
RPF-all+RBF & 0.3569 & 0.3358 & 0.3460 \\ \hline
\end{tabular}
\begin{tablenotes}
        \footnotesize
        \item \tiny 1. * means the best result; 2. RBF doesn't include reply information
\end{tablenotes}
\end{threeparttable}
\end{table} 
\vspace*{-0.2\baselineskip}

For OER recommendation task, participants rated 19.8\% of the recommended OERs as \textit{``Good''}, 23.9\% as \textit{``OK''}, 49.3\% as \textit{``Bad''}, and 6.9\% as \textit{``Not Sure''}. For this experiment, we use the globe learning-to-rank as baseline \cite{liu2015scientific} (without student clusters). From an nDCG viewpoint, we score $Good = 2, OK= 1$, and $Bad= 0$. The OER recommendation performance can be found in Table~\ref{tab:vcp}. We use nDCG@3 as the indicator to train the learning to rank models. For evaluation, 10-fold cross-validation was utilized. 

In experiment, all valid readers have RPF, we use 4-fold Cross-validation to simulate the reality situation(randomly choose 25\% students, assume they have no RPF, use the other 75\% readers to train a Max entropy classifier to predict this 25\% readers' communities).  
%For behavior feature classifier training, we use 10-fold Cross validation to simulate the reality situation(randomly choose 10\% students, assume they have no background survey information). Except baseline, we trained learning-to-rank model for each community, and averaged the recommendation performance.

As Table \ref{tab:vcp} shows, the recommendation based on RPF (\textit{p<0.001}) and RBF with Max Entropy training (\textit{p<0.05}) outperforms globe learning-to-rank (baseline) OER-recommendation method for all the metrics. The recommendation based on RPF achieved the best MAP@3 and nDCG@3, while RBF with Max Entropy training achieved  MAP@5, MAP@all, nDCG@5, nDCG@all and MRR (mean reciprocal rank).

%Based on learning-to-rank training result, we also found that the top weighted features in different community are  wide apart. For instance,  community division based on Student Reading Behavior Feature with Max Entropy training, the top feature for each community is meta-path ranking feature $W^{*}\overset{co}{\rightarrow} W\overset{s}{\rightarrow} R\overset{r}{\leftarrow} R_{slides}^{?} \overset{s}{\leftarrow} W^{*}$, $K^{*}\overset{cite}{\rightarrow} K \overset{s}{\rightarrow} R\overset{r}{\leftarrow} R_{wiki}^{?}$, $W^{*}\overset{co}{\rightarrow} W\overset{s}{\rightarrow} R\overset{r}{\leftarrow} R_{wiki}^{?} \overset{s}{\leftarrow} W^{*}$, respectively.

\vspace*{-0.75\baselineskip}\section{ANALYSIS AND CONCLUSION}

In this study, we propose a novel task to assist readers to better understand scientific publications by enabling their physical or virtual collaboration. For reader community detection, we employed two kinds of features, RPF and RBF. The former one's quality is high, but not always available. The latter one can be noisy.  

Experiment results show that: first, for physical collaboration, RPF-based clusters are more useful to predict the user physical collaboration. When RPF is not available, we can use RBF, but the performance will be compromised. Second, for virtual collaboration, the proposed two-step method can significantly improve the OER-scaffolding performance, i.e., recommend communitized OERs to help the readers from each community to better understand the scientific publications given their information needs. When RPF are partially missing, Max Entropy algorithm can be used to infer the community labels by combining partial RPF plus RBF. One limitation of this study is the lack of collaboration between different clusters. We will address this problem in the future work by exploring a soft or overlapping clustering algorithm.
%\balancecolumns
% That's all folks!
%\let\secfnt\undefined
%\newfont{\fnt}{ptmb8t at 5pt}
\vspace*{-0.8\baselineskip}
\section{Acknowledgments}
This work is supported by the Projects of National Natural Science Foundation of China (No. 61573028 and No. 61472014).
\vspace*{-0.8\baselineskip}
{
\small
\bibliographystyle{plain}
\bibliography{refs}

\begin{thebibliography}{10}

\bibitem{dennisimproving}
Alan~R Dennis, Kelly~O McNamara, Stacy Morrone, and Joshua Plaskoff.
\newblock Improving learning with etextbooks.
\newblock In {\em Proceedings of the 48th Hawaii International Conference on
  System Sciences}, pages 5253--5259, 2015.

\bibitem{gullo2008clustering}
Francesco Gullo, Giovanni Ponti, and Andrea Tagarelli.
\newblock Clustering uncertain data via k-medoids.
\newblock In {\em Scalable Uncertainty Management}, pages 229--242. Springer,
  2008.

\bibitem{johnson2010individual}
Tristan~E Johnson, Thomas~N Archibald, and Gershon Tenenbaum.
\newblock Individual and team annotation effects on students' reading
  comprehension, critical thinking, and meta-cognitive skills.
\newblock {\em Computers in human behavior}, 26(6):1496--1507, 2010.

\bibitem{liu2013generating}
Xiaozhong Liu.
\newblock Generating metadata for cyberlearning resources through information
  retrieval and meta-search.
\newblock {\em Journal of the American Society for Information Science and
  Technology}, 64(4):771--786, 2013.

\bibitem{liu2013answering}
Xiaozhong Liu and Han Jia.
\newblock Answering academic questions for education by recommending
  cyberlearning resources.
\newblock {\em Journal of the American Society for Information Science and
  Technology}, 64(8):1707--1722, 2013.

\bibitem{liu2015scientific}
Xiaozhong Liu, Zhuoren Jiang, and Liangcai Gao.
\newblock Scientific information understanding via open educational resources
  ({OER}).
\newblock In {\em Proceedings of the 38th International ACM SIGIR Conference on
  Research and Development in Information Retrieval}, pages 645--654. ACM,
  2015.

\bibitem{L2R}
Donald Metzler and W~Bruce Croft.
\newblock Linear feature-based models for information retrieval.
\newblock {\em Information Retrieval}, 10(3):257--274, 2007.

\bibitem{nigam1999using}
Kamal Nigam, John Lafferty, and Andrew McCallum.
\newblock Using maximum entropy for text classification.
\newblock In {\em IJCAI-99 workshop on machine learning for information
  filtering}, volume~1, pages 61--67, 1999.

\bibitem{novak2012educational}
Elena Novak, Rim Razzouk, and Tristan~E Johnson.
\newblock The educational use of social annotation tools in higher education: A
  literature review.
\newblock {\em The Internet and Higher Education}, 15(1):39--49, 2012.

\bibitem{pea2004social}
Roy~D Pea.
\newblock The social and technological dimensions of scaffolding and related
  theoretical concepts for learning, education, and human activity.
\newblock {\em The journal of the learning sciences}, 13(3):423--451, 2004.

\bibitem{su2010web}
Addison Su, Stephen~JH Yang, Wu-Yuin Hwang, and Jia Zhang.
\newblock A web 2.0-based collaborative annotation system for enhancing
  knowledge sharing in collaborative learning environments.
\newblock {\em Computers \&amp; Education}, 55(2):752--766, 2010.

\bibitem{Sun+11}
Y.~Sun, J.~Han, X.~Yan, P.~S. Yu, and T.~Wu.
\newblock {PathSim}: Meta path-based top-k similarity search in heterogeneous
  information networks.
\newblock In {\em Proc. 2011 Int. Conf. Very Large Data Bases (VLDB'11)},
  Seattle, WA, 2011.

\end{thebibliography}
}
\end{document}